\documentclass[10pt]{article}
\usepackage[T1]{fontenc}
\usepackage[utf8]{inputenc}

\usepackage{concmath}
\usepackage{amsmath,amsopn,amsfonts}
\usepackage{relsize}
\usepackage{array}
\usepackage{multirow}
\usepackage{graphics}
\usepackage{pdflscape}

\newcommand\ds{\displaystyle}

\makeatletter
\renewcommand\section{\@startsection {section}{1}{\z@}%
                                   {-3.5ex \@plus -1ex \@minus -.2ex}%
                                   {2.3ex \@plus.2ex}%
                                   {\Large}}
\renewcommand\subsection{\@startsection{subsection}{5}{\z@}%
                                     {-3.25ex\@plus -1ex \@minus -.2ex}%
                                     {1.5ex \@plus .2ex}%
                                     {\noindent\large\itshape}}
\makeatother

\renewcommand\d{\mathrm{d}}
\newcommand\e{\mathrm{e}}
\renewcommand\i{\mathrm{i}}

\newcommand\vk{\boldsymbol k}

\newcommand\R[1]{\mathbf R\sp{#1}}
\renewcommand\Re{{\mathrm Re}}
\newcommand\uu{\hat{\boldsymbol u}}
\newcommand\uv{\hat{\boldsymbol v}}
\newcommand\vr{\boldsymbol r}
\renewcommand\S[1]{\mathcal{S}\sp{#1}}

\newcommand\derp[3][]{\partial\sp{#1}\sb{#3}{#2}}

\newcommand\q[5]{{{#1}\sp{#2}{#3}{#4}\sp{#5}}}
\newcommand\qs[5]{\sqrt{\q{#1}{#2}{#3}{#4}{#5}}}

\newcommand\TF[1]{{\widehat{#1}}}
\newcommand\TL[1]{\MakeUppercase{#1}}
\newcommand\TFL[1]{\TF{\TL{#1}}}
\newcommand*{\nextLetter}[2][1]{%
    \edef\numLet{\expandafter\number\expandafter`#2}%
    \edef\numNewLet{\number\numexpr\numLet+#1\relax}%
    \ifnum\numNewLet>122\relax%
        \edef\numNewLet{\number\numexpr\numNewLet-26\relax}%
    \fi%
    \expandafter\char\expandafter\numNewLet%
}
\newcommand\avg\nextLetter

\DeclareMathOperator{\toL}{%
  \mathchoice%
    {\quad\mathlarger\sqsupset\quad}%
    {\,\mathlarger\sqsupset\,}%
    {\sqsupset}%
    {\sqsupset}%
}

\newcommand{\toF}[1][]{%
  \mathchoice%
    {\quad\mathlarger{\underset{#1}{\triangleright}}\quad}%
    {\,\mathlarger\triangleright\sb{#1}\,}%
    {\mathlarger\triangleright\sb{#1}}%
    {\mathlarger\triangleright\sb{#1}}%
}

\newcommand{\toFL}[1][]{%
  \mathchoice%
    {\quad\mathlarger{\underset{#1}{\sqsupset\!\triangleright}}\quad}%
    {\,\mathlarger{\sqsupset\!\triangleright}{}\sb{#1}\,}%
    {\mathlarger{\sqsupset\!\triangleright}{}\sb{#1}}%
    {\mathlarger{\sqsupset\!\triangleright}{}\sb{#1}}%
}

\title{Simultaneous double transformations of functions 
       depending on space and time} 
\author{Vincent \textsc{Rossetto}
  \footnote{e-mail:~\tt vincent.rossetto@grenoble.cnrs.fr}\\[-0.5em]
\small Laboratoire de physique et mod{\'e}lisation des milieux condens\'es,\\[-0.5em]
\small CNRS/Universit\'e Joseph Fourier\\[-0.5em]
\small Maison des Magistères, BP 166\\[-0.5em]
\small 25, avenue des Martyrs, 38042 Grenoble {\sc cedex} 9, France}
\date{\relax}

\begin{document}
\maketitle

\begin{abstract}
Performing simultaneously two transformations on functions 
of space and time (for instance a Fourier transform on the
space variable and a Laplace transform on the time variable)
is easier than performing them one after the other
when the variables are combined into particular invariant quantities. 
This is naturally also true when performing two inverse
transforms simultaneously, when the conjugated variables are combined
into a propagator. An immediate application is found in the
computation of the solutions of partial differential equations.
This article contains several general examples of such
``simultaneous double transforms'' for arbitrary analytic 
functions of space and time. 
\end{abstract}


\section*{Introduction}
Let us consider a function~$f$ depending on $r$, the distance to the origin
in $\R d$, and the time $t\in\R+$ and suppose that Fourier and Laplace
transforms of$f$ exist. 
We denote by~$k$ the Fourier conjugated variable, and by~$s$ the Laplace 
parameter. We write $f(r)\toF[d]\TF f(k)$ and $f(t)\toL\TL f(s)$
the $d$-dimensional Fourier transformation (we give a rigorous 
definition of it in Section~\ref{fourier}) and 
Laplace transformation pairs, respectively.
The Fourier-Laplace
transform of~$f$, that we denote by~$\TFL f$,
is often used to solved partial differential equations.
Performing two transforms is a demanding computation:
To solve the isotropic radiative transfer equation in two
dimensions, Sato obtained for instance an intermediate step of calculation as a
series of modified Bessel functions of the second kind and fractional order%
\cite{sato1993}.
Such expressions require using transformation handbooks like References~%
\cite{abramowitzstegun,gradsteynryzhik,ET1,PBM4} or formal computation
softwares which obfuscates the computation process. We show that 
Sato could have used the straightforward formula
\begin{equation}
\frac1{2\pi}\frac{f\left(\sqrt{t^2-r^2}\right)}{\sqrt{t^2-r^2}}
\Theta(t-r)\toFL[2]
\frac{\TL f\left(\sqrt{k^2+s^2}\right)}{\sqrt{k^2+s^2}}
\label{a}
\end{equation}
and discuss the existence and applicability of similar relations
in arbitrary dimensions.
The advantage of an expression such as~\eqref{a} is that only
one transformation appears in the final result and consequently
the analytic computation of the Fourier-Laplace transform or
inverse transform of functions of the form~\eqref{a} represent
a computational simplification. We show indeed that using~\eqref{a},
the solution of the isotropic radiative transfer in two dimension
is obtained after handling only elementary functions
and Laplace transforms.

In section~\ref{fourier} we define the Fourier transform of a function
depending on the distance~$r$ to the origin in~$\R d$
and show how it is related to the isotropic Green's function
in $d$ dimensions. In section~\ref{efros} we demonstrate the general
form of simultaneous double transformations and provide a table of 
such transformations. We illustrate the application of
formula~\eqref{a} to the isotropic radiative transfer equation
in section~\ref{rte 2d} and discuss our results.

\section{Isotropic Fourier transforms in $d$ dimensions}
\label{fourier}
The laws of ballistics state that an object starting to move at $t=0$
at constant speed~$c=1$ in the constant direction~$\uu\in\S d$
($\S d$ is the unit sphere of~$\R d$)
will find itself at the position~$\vr=t\uu$ at time~$t>0$.
We can therefore define the ballistic Green's function
\begin{equation}
g_d(\vr,\,t,\,\uu)=\delta^{(d)}\left(\vr-t\uu\right),
\label{gd}
\end{equation}
that is the distribution over space and time of the object.
The \emph{isotropic} (or radial) delta-function is the
average of~$g_d$ over all the spatial directions~$\uu$
\begin{equation}
\avg g_d(r,\,t)
  =\frac1{S_d}\int_{\S d}g_d(\vr,\,t,\,\uu) \d\uu
  =\frac{\delta(r-t)}{S_d\,r^{d-1}},
\label{G}
\end{equation}
where $S_d=2\pi^{\frac{d}2}/\Gamma(\frac{d}2)$ 
denotes the measure of the unit sphere~$\S d$ 
in~$\R d$. 
We define the Fourier transform of a function $f$ as
\begin{equation}
\TF f(\vk)=\int_{\R d} \e^{-\i\vk\cdot\vr}f(\vr)\,\d\vr.
\end{equation}
The Fourier transform of~$g_d$ is therefore
\begin{equation}
\TF g_d(\vk,\,t,\,\uu)=\exp\left(-\i t\vk\cdot\uu\right).
\label{TF gd}
\end{equation}
One observes in the expression~\eqref{TF gd} that the directional average
of~$\TF g_d(\vk,\,t,\,\uu)$ only depends on the 
product $kt$ where $k=\|\vk\|$. The integration of this expression
over~$\uu$ yields
\begin{equation}
\TF{\avg g}_d(k,\,t)=
  \frac{(2\pi)^{d/2}}{S_d}\,(kt)^{1-\frac d2}\,J_{\frac d2-1}(kt)
\label{GdJnu}
\end{equation}
where $J_\nu$ denotes the Bessel function of order~$\nu$
defined in Ref.~\cite[Eq. 9.1.20]{abramowitzstegun}.
In usual space dimensions, we have
\begin{equation}
\TF{\avg g}_1(k,x)=\cos(kx),\quad
\TF{\avg g}_2(k,x)=J_0(kx),\quad
\TF{\avg g}_3(k,x)=\frac{\sin (kx)}{kx}.
\end{equation}

An isotropic, or spherically symmetric, function~$f$ defined 
on~$\R d$ depends on~$r=\|\vr\|$ only, therefore
there exists a function~$\psi$ such that for all~$\vr\in\R d$,
$f(\vr)=\psi(\|\vr\|)$. 
We will abusively note~$f(r)$ the expression~$\psi(r)$
and observing that~$\TF f(\vk)$ also depends only on $k=\|\vk\|$,
we will abusively denote by~$\TF f(k)$ the expresion~$\chi(k)$, where
$\chi$ is the function such that
$\TF f(\vk)=\chi(\|\vk\|)$ for all~$\vk\in\R d$.
The Fourier transform of~$f$ is
\begin{equation}
\TF f(\vk)
   =\int_0^\infty f(r)\,r^{d-1}\;\d r \int_{\mathcal{S}^d}
    \e^{-\i r\vk\cdot\uu} \d\uu
   =S_d\int_0^\infty f(r)\,r^{d-1}\TF{\avg g}_d(\|\vk\|,\,r)\;\d r.
\label{TF d}
\end{equation}
The inverse Fourier
transform is obtained by a similar integral and is given by
\begin{equation}
f(r)=\frac{S_d}{(2\pi)^d}\int_0^\infty \TF f(k)\,k^{d-1}
  \TF{\avg g}_d(k,\,r)\;\d k.
\label{iTF d}
\end{equation}
The applications $r\mapsto f(r)$ and $k\mapsto \TF f(k)$ are 
related by the integrals~\eqref{TF d} and~\eqref{iTF d} which kernels depend
on the dimension~$d$. These applications form a $d$-dimensional 
isotropic Fourier pair that we denote by $f(r)\toF[d]\TF f(k)$. 

\section{Efros's composition and double transforms}
\label{efros}
The simultaneous Fourier-Laplace double transformation
is based on a Laplace pair 
\begin{enumerate}
\item[(i)] which Laplace original 
is of the form of Equation~\eqref{GdJnu}; 
\item[(ii)] that is suitable for the generalized convolution
discovered by Efros\cite{efros,novikov1984}.
\end{enumerate}
The condition~(i) requires that the Laplace original 
the Fourier transform of an isotropic delta-function~\eqref{G}.
The general form of the Laplace pair is 
\begin{equation}
\TF{\avg g}_d(k,\,\tau(t,\,u)) \toL \psi(k,\,s)\e^{-u\varphi(k,\,s)},
\label{efros pair}
\end{equation}
where~$u$ is a positive real number, $\tau(t,\,u)$ is a
real function such 
that $\tau(t,\,u)>0$ for any~$t$ and~$u$ 
and~$\varphi(k,\,s)$ is a function such that
$\Re(\varphi(k,\,s))>0$ for large enough values of $\Re(s)$.
There is no further condition on the function~$\psi(k,\,s)$.
Let me introduce here an example of a relation like
Equation~\eqref{efros pair} with $d=2$ as an illustration to this rather formal condition. This example is 
found in Ref.~\cite[Eq. 29.3.92]{abramowitzstegun}:
\begin{equation}
J_0\left(k\sqrt{t^2-u^2}\right)\toL
  \frac1{\sqrt{s^2+k^2}} \e^{-u\sqrt{s^2+k^2}}.
\label{J0Efros}
\end{equation}
Performing the Fourier inversion of the left-hand member
of Equation~\eqref{efros pair}, we obtain the simultaneous double transformation pair
\begin{equation}
 \frac{\delta\left(r-\tau(t,\,u)\right)}{S_d\, r^{d-1}}
   \toFL[d] 
   \psi(k,\,s)\e^{-u\varphi(k,\,s)}.
\label{efros double pair}
\end{equation}
Carrying on the example of Equation~\eqref{J0Efros}, 
one obtains the two-dimensional
simultaneous double transformation pair
\begin{equation}
\frac1{2\pi r}\delta\left(r-\sqrt{t^2-u^2}\right) \toFL[2]
  \frac1{\sqrt{s^2+k^2}} \e^{-u\sqrt{s^2+k^2}}.
\end{equation}

Let us now multiply the Equation~\eqref{efros double pair} by an arbitrary 
analytic function $f(u)$ and integrate over the variable~$u$.  
The left-hand side becomes a simple integration with a Dirac delta function
and the right-hand side is the Laplace transform of~$f$ for the
conjugated variable~$\varphi(k,\,s)$:
\begin{equation} 
  \frac1{S_d\,r^{d-1}}
  \int_0^\infty \delta(r-\tau(t,u)) f(u)\,\d u
  \toFL[d]
  \psi(k,\,s)\TL f\left(\varphi(k,\,s)\right).
\label{efros integrated}
\end{equation}
In our example, the function~$\tau(t,u)=\sqrt{t^2-u^2}$
takes the value~$r$ only if $t>r$ and for a unique~$u>0$ equal
to~$u_1=\sqrt{t^2-r^2}$. To compute the integral of the left-hand side of
Equation~\eqref{efros integrated} one has to take care of the integral of the
Dirac-function and replace~$\delta(r-\sqrt{t^2-u^2})$
by~$\delta(u-u_1)/|\partial_u\tau(t,\,u_1)|\Theta(t-r)
=ru_1^{-1}\, \delta(u-u_1)\Theta(t-r)$. 
We finally obtain a formula for a simultaneous double transformation
 pair of remarkable symmetry:
\begin{equation}
\frac{f\left(\sqrt{t^2-r^2}\right)}{2\pi\sqrt{t^2-r^2}}\Theta(t-r)\toFL[2]
\frac{\TL f\left(\sqrt{s^2+k^2}\right)}{\sqrt{s^2+k^2}}.
\label{2DSDT}
\end{equation}
The simultaneous double transformation reduces in Equation~\eqref{2DSDT} to a
\emph{single} Laplace transformation.

In the general case, we have to 
consider, for fixed~$r$ and $t$, the equation \( r=\tau(t,\,u) \)
for the unknown~$u$ 
and call~$u_n(r,t)$, $n=1,\,2,\,\dots,\,N(r,t)$ the solutions
of the equation. 
The general result follows 
\begin{equation}
\frac{1}{S_d\, r^{d-1}}\sum_{n=1}^{N(r,t)} 
\frac{f\left(u_n(r,t)\right)}
  {\left|\derp\tau u(t,u_n(r,t))\right|}
    \toFL[d] \psi(k,\,s)\;\TL f\left(\varphi(k,\,s)\right).
\label{Efros}
\end{equation}
Note that in the expression~\eqref{2DSDT}, the $\Theta$ function
translates the fact that $N(r,t)=1$ if $r<t$ and $N(r,t)=0$ otherwise.
Examples of Fourier-Laplace pairs obtained using Efros's composition
theorem are presented in Table~\ref{table}. 

\section{Application to the isotropic radiative transfer}
\label{rte 2d}
The radiative transfer equation in two dimensions has been solved in the
isotropic case by Sato and Paasschens\cite{sato1993,paasschens1997}.
The latter having also provided the directional elementary solution.
The equation reads
\begin{equation}
 \derp jt(\vr,\,t,\,\uu)-c\uu\cdot\nabla_{\vr} j(\vr,\,t,\,\uu)+\frac c\ell
   j(\vr,\,t,\,\uu)
  =\frac c\ell\int_{\S2} j(\vr,\,t,\,\uv) \d\uv 
\label{RTE}
\end{equation}
where $c$ is the wave celerity, $\ell$ the extinction length, $\uu$
is a unit vector and the function~$j(\vr,t,\uu)$ is the radiance:
The energy flowing per time unit 
through an area element~$\d A$ with normal vector~$\uu$
at point $\vr$ and time~$t$ is $j(\vr,\,t,\,\uu)\d A$. 
We denote by~$i(r,\,t)$ the
directional average of~$j(\vr,\,t,\,\uu)$. 
The Fourier-Laplace transform of the equation~\eqref{RTE} is
\begin{equation}
\TFL\jmath(\vk,\,s,\,\uu)-\TF\jmath(\vk,\,0,\,\uu)
  +\i c\uu\cdot\vk\TFL j(\vk,\,s,\,\uu)
  +\frac c\ell\TFL j(\vk,\,s,\,\uu)=\frac c\ell\int_{\S2}\TFL j(\vk,\,s,\,\uv)
  \d\uv
\end{equation}
Recognizing in Equation~\eqref{gd} the expression 
of~$\TFL g_2(k,\,s,\,\uu)=(s+\i c\vk\cdot\uu)^{-1}$,
the equation rewrites after the 2-D Fourier-Laplace simultaneous 
double transformation
\begin{equation}
  \TFL j(\vk,\,s,\,\uu)=
  \TFL g_2\left(\vk,\,s+\frac c\ell,\,\uu\right)
    \TF\jmath(\vk,\,0,\,\uu)
  +\frac c\ell\,\TFL g_2\left(\vk,\,s+\frac c\ell,\,\uu\right)
   \TFL i(k,\,s), 
\label{Dyson}
\end{equation}
and taking the directional average with the isotropic initial condition
$j(\vr,\,0,\,\uu)=\frac{A_0}{2\pi}\delta^{(2)}(\vr)$, we get the expression
\begin{equation}
\TFL i(k,\,s)=A_0\frac{\TFL{\avg g}_2\left(k,\,s+\frac c\ell\right)}
{1-\frac c\ell\TFL{\avg g}_2\left(k,\,s+\frac c\ell\right)}.
\label{I(H2)}
\end{equation}
Let us use the equality~$\TFL{\avg g}_2(k,\,s)=(s^2+k^2)^{-1/2}$ 
and formula~\eqref{2DSDT} 
(also given as entry \ref{t2-r2} of Table~\ref{table} for $d=2$) 
with the function~$f$
such that~$\TL f(s)=s/(s-c/\ell)$. The inversion
of~$\TL f$ is elementary and yields
$f(t)=\delta(t)+(c/\ell)\exp[ct/\ell]$. We finally obtain the
result, as found in the 
References~\cite{sato1993,paasschens1997},
\begin{equation}
  i(r,\,t)
   =\frac{A_0}{2\pi}\left[\frac{\delta(r-ct)}{r}
     +\frac{\e^{\frac1\ell\sqrt{c^2t^2-r^2}}}
     {\ell\sqrt{c^2t^2-r^2}}
     \Theta(ct-r)\right]\e^{-ct/\ell}.
\label{J}
\end{equation}
The leading decreasing exponential is given by an elementary property
of the Laplace transform.
Note that the delta function requires a particular attention
to get the final result. The directional solution~$j(\vr,\,t,\,\uu)$
is obtained using the expression~\eqref{J} in the Equation~\eqref{Dyson}.
We have thus obtained the solution~\eqref{J} by means of the general form of the
2-D Fourier-Laplace simultaneous double transformation\eqref{2DSDT}
and some basic properties of the elementary function~$\avg g_2$.

\section{Discussion}
\label{discussion}
The simultaneous inversion of the space-Fourier and the time-Laplace
transformations can be performed in certain particular cases, when 
the conjugated variables are combined in a simple way
in the Fourier-Laplace domain. In these cases, only the computation of a
single inversion is required to obtain the double inverse in the
space-time domain. 
Table~\ref{table} displays a summary of the results
presented in this article. Looking up at this table one can remark
that in all of its entries the variables are combined
in a similar way in the space-time domain and in the Fourier-Laplace
domain. Such symmetry between linear combinations in space-time
domain and their \emph{conjugated combinations} in the Fourier-Laplace
follows from the similarity between the Fourier
transformation, the Laplace transformation and their inverse
transformations. They merely reflect the fact that using an 
image as an original of another transformation is almost
like performing an inverse transformation on this image.
But we have not only obtained relations between functions
of linear combinations:
Consider for instance the entry~\ref{t2-r2}
in Table~\ref{table}, where the variable combination
is $Q(r,t)=\qs t2-r2$ in the space-time domain and $\qs s2+k2$
in the conjugated space, or the entry~\ref{r^2/t}, where 
these combinations are respectively~$r^2/4t$ and~$k^2/s$. 
These simultaneous double transformation pairs have been obtained using
intermediate transforms that do not display such
symmetry, it appears therefore that the simultaneous double transformations
preserve, in certain cases, more symmetry than single transformations.

Some combinations of space and time that appear in the Table~\ref{table}
are invariants for a physical phenomenon. The combination
$t-r$ in the entries \ref{(t-r)/r}, \ref{(t-r)/r^(d/2)} 
and \ref{(t-r)/r^(d/2+1)} is invariant under galilean transformations,
the combination $r^2/t$ reflects the self-similarity of diffusion,
and $Q=\sqrt{t^2-r^2}$ is a proper time in special relativity.
Reciprocally, the conjugated combinations are closely related
to the propagators of these invariants. For instance
in the application to the radiative transfer described in Section~\ref{rte 2d}
the solution~\eqref{I(H2)} is expressed thanks to 
the free propagator~$\TFL{\avg g}_2$. The entry~\ref{t2-r2}
for~$d=2$ can be rewritten as
$f(Q)/2\pi Q\toFL[2]\TFL{\avg g}_2\TL f(\TFL{\avg g}_2^{-1})$.
We may interprete this simultaneous double transformation as
a description in terms of free propagators in the same way as the Fourier
transform is a description in terms of modes.

The combinations present in Table~\ref{table} are the only ones 
obeying the requirements of the composition summarized in 
Equation~\eqref{efros pair} that can be derived from the reference tables 
\cite{abramowitzstegun,ET1,PBM4}.
They have been discovered and used 
because of their frequent appearance in the equations of Physics.
One may nevertheless wonder whether there are not other 
Fourier-Laplace simultaneous double transformation pairs for these, or other,
combinations (invariants or propagators).
Furthermore, the question of finding new simultaneous double transformation
pairs extends to any other double transformation, replacing the spatial
Fourier transformation or the time Laplace transformation by 
other transformations taken
among the large set of indexed integral transformations 
(Mellin transform, $z$-transform, Bessel transforms\dots). 

The general principle of simultaneous double transformation can be extended to
higher dimensions and other transforms. It may prove useful in other
physical sciences where partial differential equations are important.
Double transformations will hopefully help solving more partial differential 
equations in an easier way. 

\section*{Acknoledgments}
This work was funded by the
regular annual endowments
of the Centre National de la Recherche Scientifique and of 
the Universit{\'e} Joseph Fourier
to the Laboratoire de Physique et Mod{\'e}lisation des Milieux 
Condens{\'e}s.


\begin{thebibliography}{1}

\bibitem{abramowitzstegun}
{\sc Milton Abramowitz and Irene~A. Stegun}, {\em Handbook of mathematical
  functions}, Dover, 10th~ed., 1972.

\bibitem{efros}
{\sc A.~M. Efros}, {\em O nekotorikh primeneniakh operatornogo itsislenia
  analisu}, Matematicheskii Sbornik, 42 (1935), pp.~699--705.

\bibitem{ET1}
{\sc A.~Erd{\'e}lyi}, {\em Handbook of integral transforms, volume I}, McGraw
  Hill, New-York, 1954.

\bibitem{gradsteynryzhik}
{\sc I.~S. Gradsteyn and I.~M. Rhyzhik}, {\em Table of integrals, series and
  products}, Academic Press, London, sixth edition~ed., 2000.

\bibitem{novikov1984}
{\sc I.~A. Novikov}, {\em Change of variables in the {L}aplace transform and
  some applications}, Journal of engineering physics, 47 (1984),
  pp.~1103--1109.

\bibitem{paasschens1997}
{\sc J.~C.~J. Paasschens}, {\em Solution ot the time-dependent {B}oltzmann
  equation}, Phys. Rev. E, 56 (1997), pp.~1135--1141.

\bibitem{PBM4}
{\sc A.~P Prudnikov, Yu.~A. Brychkov, and O.~I. Marichev}, {\em Integrals and
  series, vol. 4 direct Laplace transforms}, Gordon and Breach Science
  Publishers, New-York, 1990.

\bibitem{sato1993}
{\sc Haruo Sato}, {\em Energy transport in one- and two-dimensional scattering
  media: Analytic solutions of the multiple isotropic scattering model}, Geo.
  Phys. J. Int., 117 (1993), pp.~487--494.

\end{thebibliography}

\newcounter{type}
\newcounter{pair}[type]

\makeatletter
\newcommand\tablecite[2][]{\hbox{Ref.~\cite{#2}, eq.~{#1}}}
\renewcommand\thepair{%
  \thetype.\@arabic\c@pair }
\makeatother
\setlength\baselineskip{20pt}
\refstepcounter{type}
\setcounter{pair}{0}

\begin{landscape}
\begin{table}
\caption{\label{table}
Table of simultaneous Fourier-Laplace double transformations
for spherically symmetric function in space dimension~$d$. 
These transforms are based on the Equation~\eqref{Efros}
applied to several identities givern in reference.
The first index distinguishes between 1) simple multiplication of
$\TL f(s)$ and 2) more complicated arguments of~$\TL f$.
The dots at the end of some lines indicate that the expression
extends to the next line.}
\begin{tabular}{>{\refstepcounter{pair}\thepair}cr@{}lr@{}lm{3.5cm}}
\hline
  \multicolumn{1}{c}{\#}& 
  \multicolumn{2}{c}{\bf Fourier-Laplace domain} & 
  \multicolumn{2}{c}{\bf Space-time domain} & References and notes \\
\hline
\multicolumn{1}{c}{}&&&&&\\[-0.5em]
\label{(t-r)/r} & 
$\ds (s^2+k^2)^{\frac{1-d}2}$ & 
$\TL f(s)$& 
$\ds \frac{\pi S_{d-1}}{(2\pi)^d}\,\frac1r$ & 
$f(t-r)\Theta(t-r)$ &
\hbox{$d\geq 2$}
\tablecite[1.1.5.30]{PBM4}\\
\label{(t-r)/r^(d/2)} & 
$\ds \frac{(s+\qs s2+k2)\sp{1-d/2}}{\qs s2+k2}$ & 
$\TL f(s)$& 
$\ds \frac{1}{(2\pi r)\sp{d/2}}$ & 
$f(t-r)\Theta(t-r)$ &
\tablecite[1.1.5.28]{PBM4}\\
\label{(t-r)/r^(d/2+1)} & 
$\ds \left(s+\qs s2+k2\right)^{1-d/2}$ & 
$\TL f(s)$& 
$\ds \frac{\frac d2-1}{(2\pi)^{d/2}}\,\frac1{r^{\frac d2+1}}$ & 
$f(t-r)\Theta(t-r)$ &
\hbox{$d\neq2$}
\tablecite[1.1.5.29]{PBM4}\\
\label{t-r2} & 
$\ds \frac1{s\sp{d/2}}\e\sp{-k\sp{2}/4s}$ & 
$\TL f(s)$ & 
$\ds \frac1{\pi\sp{d/2}}$ & 
$f(t-r\sp2)\Theta(t-r\sp2)$ &
\tablecite[1.1.5.32]{PBM4}\\
\label{t+a-sqrt(t^2-a^2)} & 
$\exp\left[-a(\qs s2+k2\,-s)\right]\dots$ & 
& 
$\ds\frac{1}{(2\pi)\sp{d/2}}\frac{(a+\qs r2+a2)\sp{1-d/2}}{\qs r2+a2}$ &
$    f\left(t+a-\qs r2+a2\right)\dots$ & 
\\
\multicolumn{1}{c}{}& 
$  \qquad\times\ds\frac{(s+\qs s2+k2)\sp{1-d/2}}{\qs s2+k2}$ &
$\TL f(s)$ &
&
$ \qquad\times\Theta\left(t+a-\qs r2+a2\right)$ &
\tablecite[1.1.5.31]{PBM4}
\refstepcounter{type}
\setcounter{pair}{0}
\\
\hline
\multicolumn{1}{c}{}&&&&&\\[-0.5em]
\label{t2-r2} & 
$\ds \frac{(s+\qs s2+k2)\sp{1-d/2}}{\qs s2+k2}$ &
$\ds \TL f\left(\qs s2+k2\right)$ &
$\ds \frac{1}{(2\pi)\sp{d/2}}\frac{(t+\qs t2-r2)\sp{1-d/2}}{\qs t2-r2}$ &
$\ds f\left(\qs t2-r2\right)\Theta(t-r)$ &
\tablecite[29.3.97]{abramowitzstegun}
\tablecite[1.1.5.34]{PBM4}\\
\label{r^2/t}& 
$\ds \frac{1}{s\sp{d/2}}$ &
$\ds \TL f\left(\frac{k\sp2}s\right)$ &
$\ds \frac{1}{(2\pi)\sp{d/2}} \frac{r\sp{2-d}}{(2t)\sp{2-d/2}}$ & 
$\ds f\left(\frac{r\sp2}{4t}\right)$ &
\tablecite[29.3.80]{abramowitzstegun}
\tablecite[1.1.5.27]{PBM4}\\
\label{(r2-t2)/t} & 
$\ds \frac{(s+\qs s2+k2)\sp{1-d/2}}{\qs s2+k2}$ & 
$\ds \TL f\left(\qs s2+k2\,-s\right)$ & 
$\ds \frac{1}{(2\pi)\sp{d/2}} \frac{r\sp{2-d}}{t\sp{2-d/2}}$ &
$\ds f\left(\frac{\q r2-t2}{2t}\right) \Theta(r-t)$ & 
\tablecite[4.15(21)]{ET1}\\
\label{t-sqrt(t^2-r^2)}& 
$\ds \frac{1}{s\sp{d/2}}$ & 
$\ds \TL f\left(s+\frac{k\sp2}{4s}\right)$ &
$\ds \frac{1}{(2\pi)\sp{d/2}} \frac{1}{t+\qs t2-r2}$ &
$\ds f\left(t-\qs t2-r2\right)\Theta(t-r)$ &
\tablecite[1.1.5.38]{PBM4}\\
\hline
\end{tabular}
\end{table}
\end{landscape}

\end{document}